\documentclass{article}

\usepackage{arxiv}
\usepackage{enumitem}
\usepackage{graphicx}
\usepackage[utf8]{inputenc} 
\usepackage[T1]{fontenc}    
\usepackage{hyperref}       
\usepackage{url}            
\usepackage{booktabs}       
\usepackage{amsfonts}       
\usepackage{nicefrac}       
\usepackage{microtype}      
\usepackage{lipsum}
\usepackage{subcaption}
\usepackage[lined,plainruled,linesnumbered]{algorithm2e}

\title{Driven tabu search: a quantum inherent optimisation}

\author{
  Carla Silva and Inês Dutra \\
  Department of Computer Science\\
  Faculty of Sciences, University of Porto\\
  Porto, Portugal \\
  \texttt{carla.maps@gmail.com}, \texttt{ines@dcc.fc.up.pt} \\
   \And
 Marcus S. Dahlem \\
  Department of Electrical Engineering and Computer Science\\
  Masdar Institute, Khalifa University of Science and Technology\\
  Abu Dhabi, United Arab Emirates \\
  \texttt{marcus.dahlem@ku.ac.ae} \\
}

\begin{document}
\maketitle

\begin{abstract}
Quantum computers are different from binary digital electronic computers based on transistors. Common digital computing encodes the data into binary digits (bits), each of which is always in one of two definite states (0 or 1), quantum computation uses quantum bits (qubits). A circuit-based qubit quantum computer exists and is available for experiments via cloud, the IBM quantum experience project. We implemented a Quantum Tabu Search in order to obtain a quantum combinatorial optimisation, suggesting that an entanglement-metaheuristic can display optimal solutions and accelerate the optimisation process by using entangled states. We show by building optimal coupling maps that the distribution of our results gave similar shape as shown previous results in an existing teleport circuit. Our research aims to find which graph of coupling better matches a quantum circuit.
\end{abstract}


\section{Introduction}
In Quantum Mechanics, a qubit is a quantum system in which the Boolean states 0 and 1 are represented by a pair of normalised and mutually orthogonal quantum states. The two states form a computational basis and any other state of the qubit can be written as a linear combination of $|0\rangle$ and $|1\rangle$: $|\psi\rangle = \alpha|0\rangle + \beta|1\rangle$ , where $\alpha$ and $\beta$ are probability amplitudes and can be complex numbers. A qubit can be in a superposition of both states at the same time. Multiple qubits can exhibit quantum entanglement - pairs are generated or interact in ways that the quantum state of each particle cannot be described independently of the state of the other(s) ~\cite{Wittek2014}. The IBM (International Business Machines Corporation) Quantum Experience (QX) allows us the possibility to connect to an IBM quantum processor via the IBM Cloud. We developed a driven quantum version of the Tabu search algorithm ~\cite{Glover1986,Glover1997}, which has been well understood for solving combinatorial or nonlinear problems ~\cite{Kuo2017}. The experiments were implemented in the Python programming language using the Quantum Information Software Kit (QISKit) - a software development kit (SDK) for working with the Open Quantum Assembly Language (OpenQASM) and the IBM QX. We use as backend IBM Q 16 Rueschlikon (16-qubits) and IBM Q 5 Yorktown (5-qubits) simulators. In a quantum-metaheuristic procedure, we challenge the quantum search space to be maximized, derived from the advantage of interacting quantum technologies with classical implementations. In a quantum combinatorial optimisation, an entanglement-metaheurisc can uncover optimal solutions and accelerate the optimisation process by using entangled states. Therefore, we conduct simulation-based experiments with two types of quantum initial populations. The sample solutions are composed by 16 qubits of combined pairs - with non replacement, and with replacement. Until we obtain the best solution, we build the neighborhood, evaluate the system and detect whether the algorithm falls in a local optimum. In those cases, we perform the entanglement between qubits if they are unequal; otherwise, we use superposition in the redundant qubit. This particular feature of enhanced-entanglement showed different best solutions, and different system evaluation according to different combinatorial inputs. The quantum inherent optimisation allows us to highlight the best solution and algorithm performance according to the input combined set. We show the application of the proposed research by defining coupling maps for quantum devices with a driven Tabu search approach.

\section{Quantum computing}
\label{sec:concepts:sec1}

Quantum programming is like composing, where qubits in superposition interfere and where the quantum programmer ensures useful interference of qubit states.  
Experimental measurements disturbs, or can destroy, the wave-like quantum states. In order to better understand such states through experiments, it is possible to gather detailed properties of the quantum system by averaging many weak nondisturbing quantum measurements. However, the frontier between what is quantum and what is classical becomes small when averaging over a large number of weak measurements.

Quantum mechanics allows the qubit to be in a superposition of both states at the same time. In quantum computing, a qbit or qubit or quantum bits a unit of quantum information. Therefore, a qubit is a two-state quantum-mechanical system, similar to the polarisation of a single photon, where the two states are vertical polarisation and horizontal polarisation, since it can be described as a polarised photon. A quantum register of size \textit{n} is defined by \textit{n} qubits. A qubit is a quantum system in which the Boolean states 0 and 1 are represented by a pair of normalised and mutually orthogonal quantum states. The two states form a computational basis and any other (pure) state of the qubit can be written as a linear combination of $|0\rangle$ and $|1\rangle$: $|\psi\rangle = \alpha|0\rangle + \beta|1\rangle$ , where $\alpha$ and $\beta$ are probability amplitudes and can be complex numbers. For example, $|\psi\rangle = \frac{1}{\sqrt[]{3}}|0\rangle + \sqrt[]{\frac{2}{3}}|1\rangle$, with probabilities $|\alpha_0|^2 = \frac{1}{3}$ and $|\alpha_1|^2 = \frac{2}{3}$, and measurement result 0 and 1, respectively.

In quantum mechanics, bra-ket notation is a standard notation for representing quantum states. In order to calculate the scalar product of vectors, the notation uses angle brackets $\langle$ $\rangle$, and a vertical bar |. The scalar product is then $\langle\phi|\psi\rangle$ where the right part is the "psi ket" (a column vector) and the left part is the bra - the Hermitian transpose of the ket (a row vector). 

\begin{table}[h]
\small
  \centering
  \begin{tabular}{ccc}
    Qubits & Classical States & Superposition\\
    \hline
     &  & $\alpha_0|0\rangle + \alpha_1|1\rangle$\\
    1&2&$|\alpha|^2 + |\beta|^2 = 1$\\
    \hline
     &  & $\alpha_{00}|00\rangle + \alpha_{01}|01\rangle + \alpha_{10}|10\rangle + \alpha_{11}|11\rangle$\\
    2&4&$|\alpha_{00}|^2 + |\alpha_{01}|^2 + |\alpha_{10}|^2 + |\alpha_{11}|^2 = 1$\\
    \hline
     &  & $\sum_{x=0}^{2^n-1} \alpha_x |x\rangle$ \\
    $n$&$2^n$&$\sum_{x=0}^{2^n-1} |\alpha_x|^2 = 1$\\
    \hline
  \end{tabular}
  \caption{Exponential state space. Where $|\alpha_i|^2$ is the probability of finding the qubit in state $|i\rangle$ when we measure it (in the computational basis).}
  \label{tab:qubits}
\end{table}

\subsection{Superposition}
\label{sec:concepts:sec2}

Since a pure qubit state is a linear superposition of the basis states when we measure the qubit, according to the Born rule, the probability of outcome $|0\rangle$ is $|\sqrt[2]{\alpha|}$ and the probability of outcome is $|1\rangle$ is $|\beta|^2$, $\alpha$ and $\beta$ are constrained by the equation $|\alpha|^2 + |\beta|^2$. Superposition is then similar to waves in classical physics, any two (or more) quantum states can be added, becoming superposed resulting in a new quantum state - i.e. in 0 and 1 simultaneously, a linear combination of the two classical states - for example, the states $|+\rangle ={\frac  {1}{{\sqrt  {2}}}}(|0\rangle +|1\rangle )$ or $|-\rangle ={\frac  {1}{{\sqrt{2}}}}(|0\rangle -|1\rangle )$.

\subsection{Entanglement}
\label{sec:concepts:sec3}

An important difference between a qubit and a classical bit is that multiple qubits can exhibit quantum entanglement - a physical phenomenon that occurs when pairs or more than two particles are generated or interact in forms that the quantum state of each particle can't be described independently of the state of the other(s). In the case of two entangled qubits in the Bell state $\frac{1}{\sqrt[]{2}}(|00\rangle + |11\rangle)$, an equal superposition, there are equal probabilities of measuring either $|00\rangle$ or $|11\rangle$, as $|\frac{1}{\sqrt[]{2}}|^2 = \frac{1}{2}$. Considering the two qubits are entangled qubits separately, Alice's qubit and Bob's qubit. Alice measure her qubit, obtaining equal probabilities in either $|0\rangle$ or $|1\rangle$. Since the qubits are entangled, Bob must now get exactly the same measurement as Alice. 

\section{Quantum Tabu Search}
\label{sec:headings}
Some perspectives of Quantum Tabu Search (QTS) have been presented ~\cite{ChiangCCKH14,WangCKHC12}. The classical Tabu Search (TS) is a metaheuristic that explores search spaces and conduct a local heuristic search procedure to explore the solution space beyond local optimum using a Tabu list with forbidden moves. The algorithm~\ref{alg} main steps can be briefly described by, a) generating the neighbors, b) evaluating each neighbor and c) getting the neighbor with maximum evaluation. The algorithm stops at any iteration where there are no feasible moves into the local neighborhood of the current solution.

The Quantum Tabu Search is proposed for solving combinatorial or nonlinear problems with a Knapsack problem approach in a quantum procedure: qubit ($i$), weight ($w_i$) and profit ($b_i$),

\begin{equation}
f(s)=\sum_{i=1}^{n}b_is_i*\left(1-max\left(0,\sum_{i=1}^{n}w_is_i-max capacity\right)\right)
\end{equation} where $s$ is the best solution and the goal is to achieve a global optimum by maximizing the quantum parameters. 

Coupling maps for quantum devices, can be an application of the QTS, where qubit–qubit couplings are explored based on combinations of qubits taking into account the maximum number in a given quantum architecture.

\begin{figure}[ht]
  \centering
  \includegraphics[height=3cm]{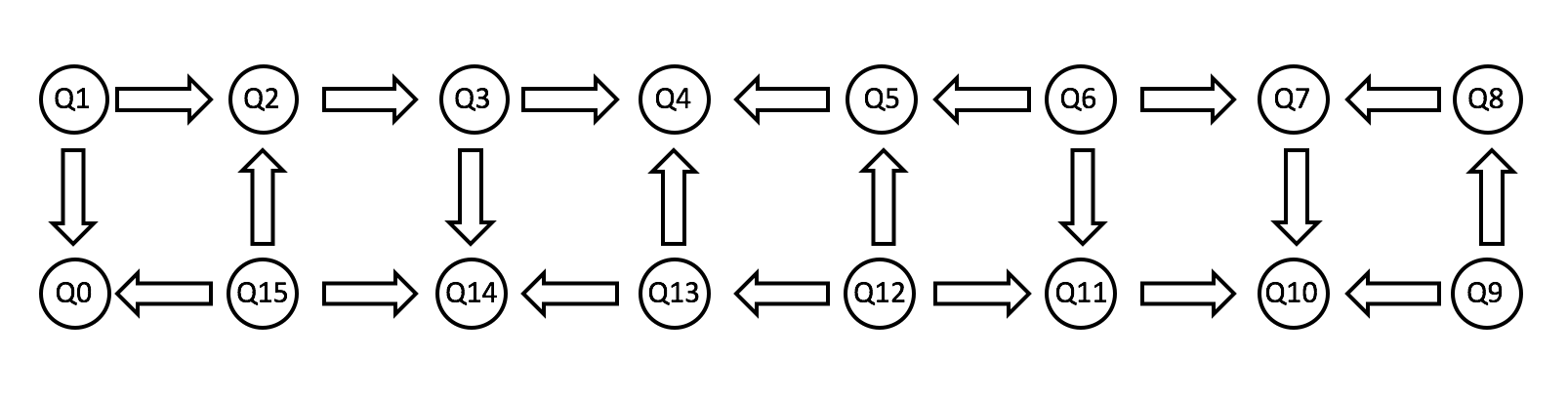}
  \caption{IBMQX5 connections, 16-qubits.}
  \label{fig:fig1}
\end{figure}

A coupling map in quantum architectures is a directed graph representing superconducting bus connections between qubits~\cite{2018arXiv} as shown in Figure~\ref{fig:fig1}.

To obtain the neighbor with maximum evaluation we get the tabu qubit, find the best neighbor position, afterwards we check if the neighbor is a result of a tabu move, if it is, then we get the position of the qubit and check if it is in a tabu list. The tabu list is built by storing the qubit considering if the neighbors evaluation is greater that best evaluation. The \textit{best iteration} is an incremented value when the condition occurs and the \textit{best solution} are the neighbors given by the position of best neighborhood evaluation. The qubit is given by the position that occurs in the range of the length of the best solution when the best solution differs from the best neighbor. The input and output states will be in a superposition, so that the qubits can be also entangled. 

\begin{figure}[ht]
  \centering
  \begin{minipage}{.5\linewidth}
  \centering
\begin{algorithm}[H]             
\DontPrintSemicolon
\KwResult{best solution, evaluation, best iteration}
 initialize quantum population $Q(t)$\;
 \While{True}{
  \If{$(iteration - best iteration) > max quantity$}{
      \eIf{$best solution[0] \neq best solution[1]$}{
          apply CNOT gate\;
      }{
          apply H gate\;
      }
   }
   new neighborhood generation and evaluation\;
   \If{$neighborsevaluation > bestevaluation$}{
	keep tabu move\;
	new better solution $s^b$\;
	update $Q(t)$ best evaluation $e^b$\;
    increment best iteration $i^b$\;
   }
 }
 \caption{General QTS algorithm} \label{alg}
\end{algorithm}
\end{minipage}
\end{figure}

\section{Results: find the best coupling map}
\label{sec:Results}

\begin{figure}[ht]
    \centering
    \begin{subfigure}[t]{0.5\textwidth}
        \centering
        \includegraphics[width=\columnwidth]{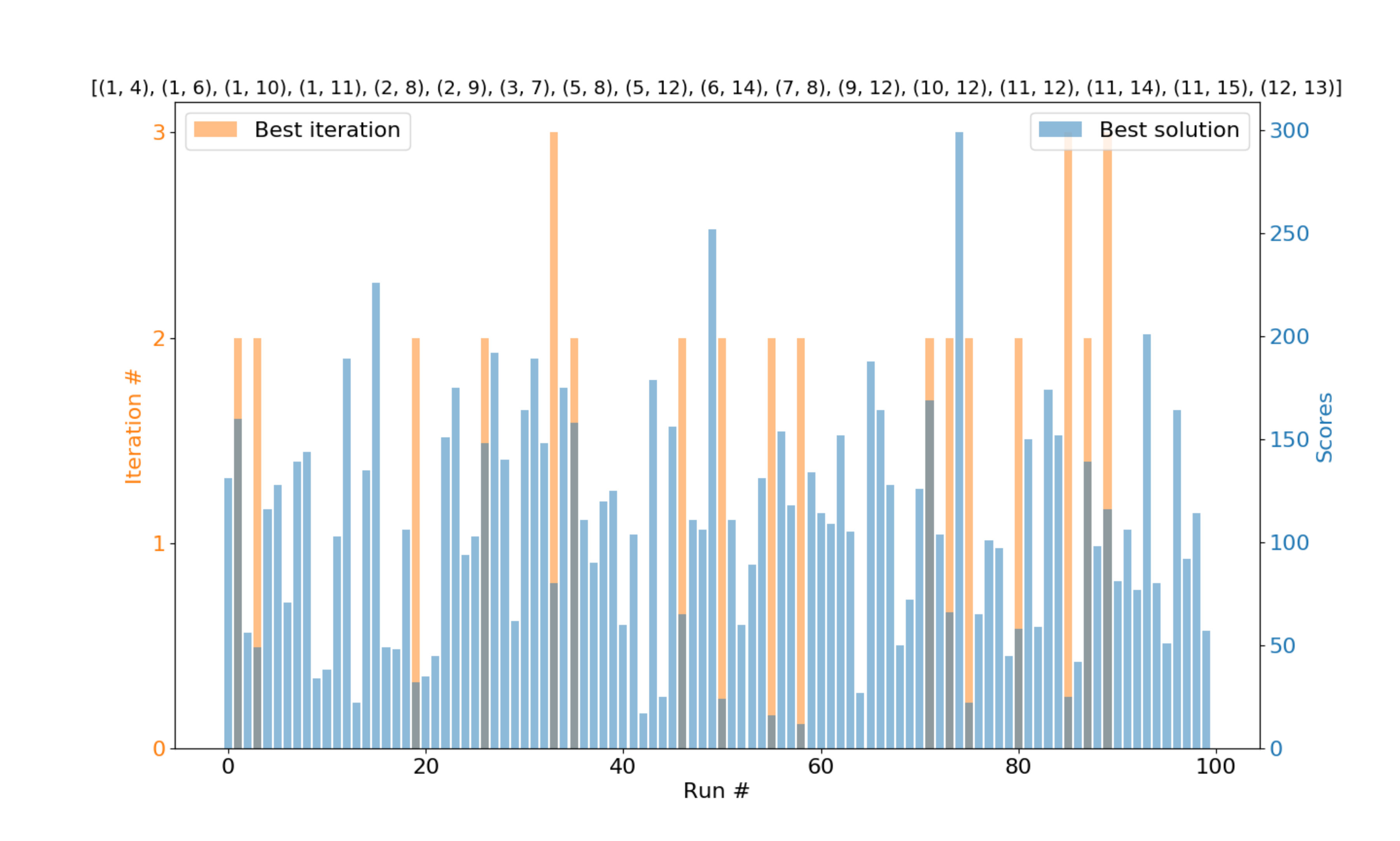}
        \caption{Many pairs combination.}\label{fig:cmap1}
    \end{subfigure}%
    ~ 
    \begin{subfigure}[t]{0.5\textwidth}
        \centering
        \includegraphics[width=\columnwidth]{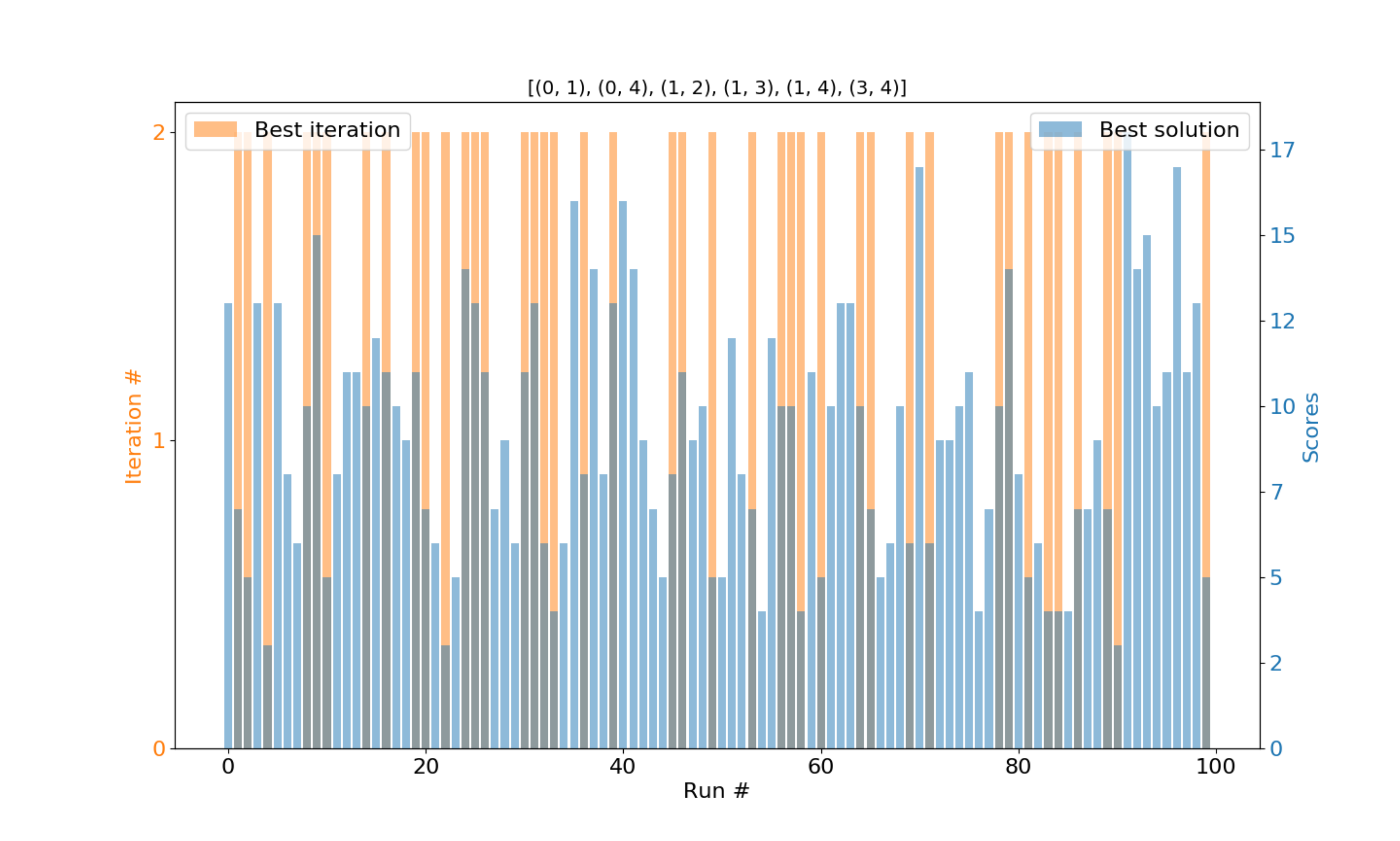}
        \caption{Few pairs combination.}\label{fig:cmap2}
    \end{subfigure}
    \caption{100 runs to find best coupling map.}
\end{figure}

In Figure~\ref{fig:cmap1} and Figure~\ref{fig:cmap2} we present two combinatorial configurations of pairs in 100 runs each. The figures show the scores for each solution and the respectively number of iterations. A maximum number of iterations do not particularly mean a high score solution.

To validate our results we choose the quantum teleport circuit\footnote{https://github.com/Qiskit/qiskit-terra} with parameters initialization and quantum gates performing operations, such us, Pauli-$\it{X}$ (NOT Gate) to obtain a $\pi$-rotation around the $\it{X}$ axis where $\it{X}\rightarrow\it{X}$ and $\it{Z}\rightarrow-\it{Z}$ (bit-flip), Pauli-$\it{Z}$ ($\it{R}_\pi gate)$) to obtain $\pi$-rotation around the $\it{Z}$ axis where $\it{X}\rightarrow-\it{X}$ and $\it{Z}\rightarrow\it{Z}$ (phase-flip), Hadamard $\it{H}$ to map $\it{X}\rightarrow\it{Z}$, and Controlled-NOT, a two-qubit gate that flips the target qubit (applies Pauli-$\it{X}$). 

\begin{figure}[ht]
\centering
\includegraphics[width=0.5\linewidth,height=3cm]{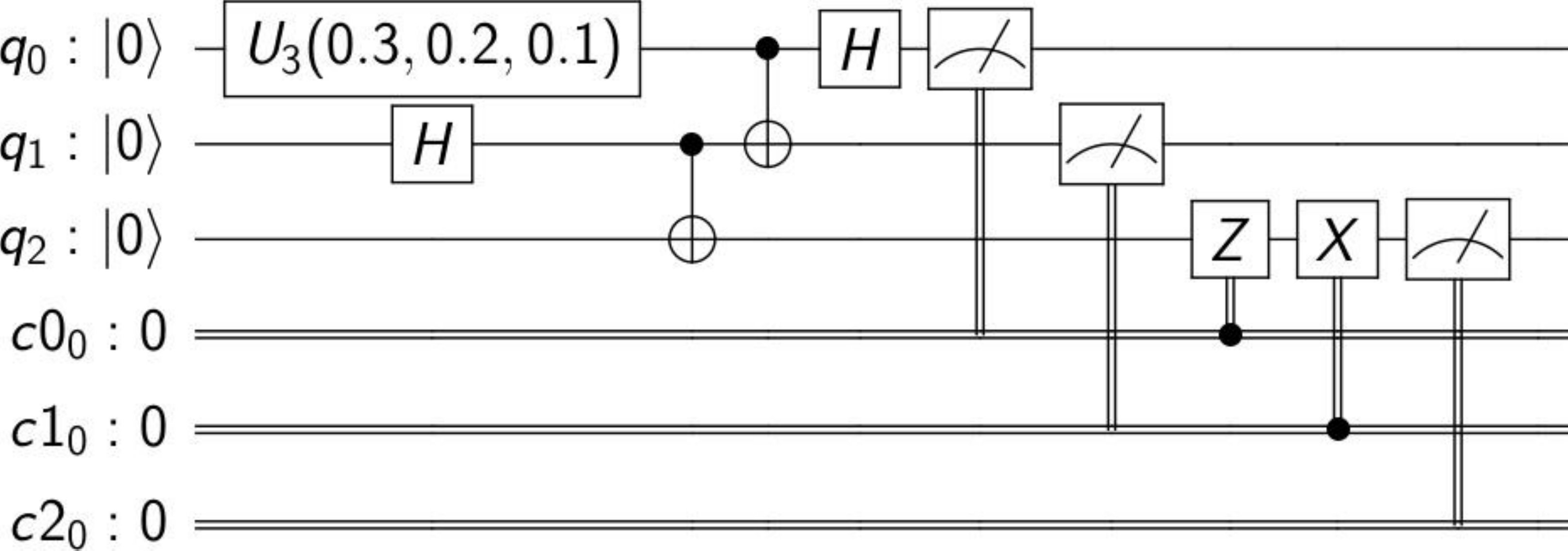}
\caption{Quantum teleport circuit.} \label{fig:2}
\end{figure}

\begin{enumerate}[label=\alph*]
\item Coupling map = \textit{None}
\item Coupling map = [[0, 1], [0, 2], [1, 2], [3, 2], [3, 4], [4, 2]]
\item Coupling map = [[0, 1], [0, 4], [1, 2], [1, 3], [1, 4], [3, 4]] (Tabu)
\end{enumerate}
\begin{figure}
\centering
\begin{subfigure}{0.33\textwidth}
\centering
\includegraphics[width=\linewidth,height=3cm]{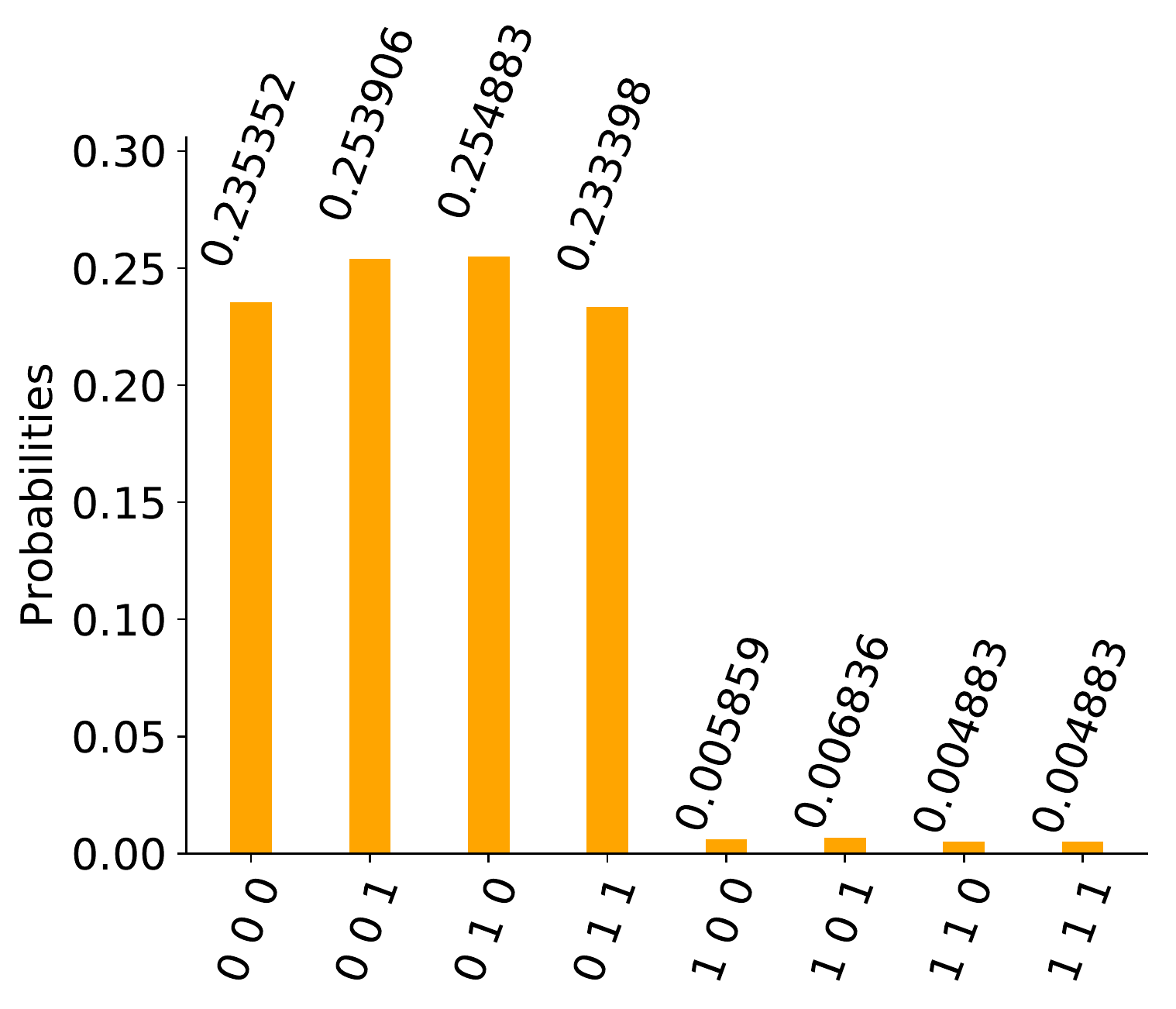}
\caption{Coupling map 1.} \label{fig:4}
\end{subfigure}\hspace*{\fill}
\begin{subfigure}{0.33\textwidth}
\centering
\includegraphics[width=\linewidth,height=3cm]{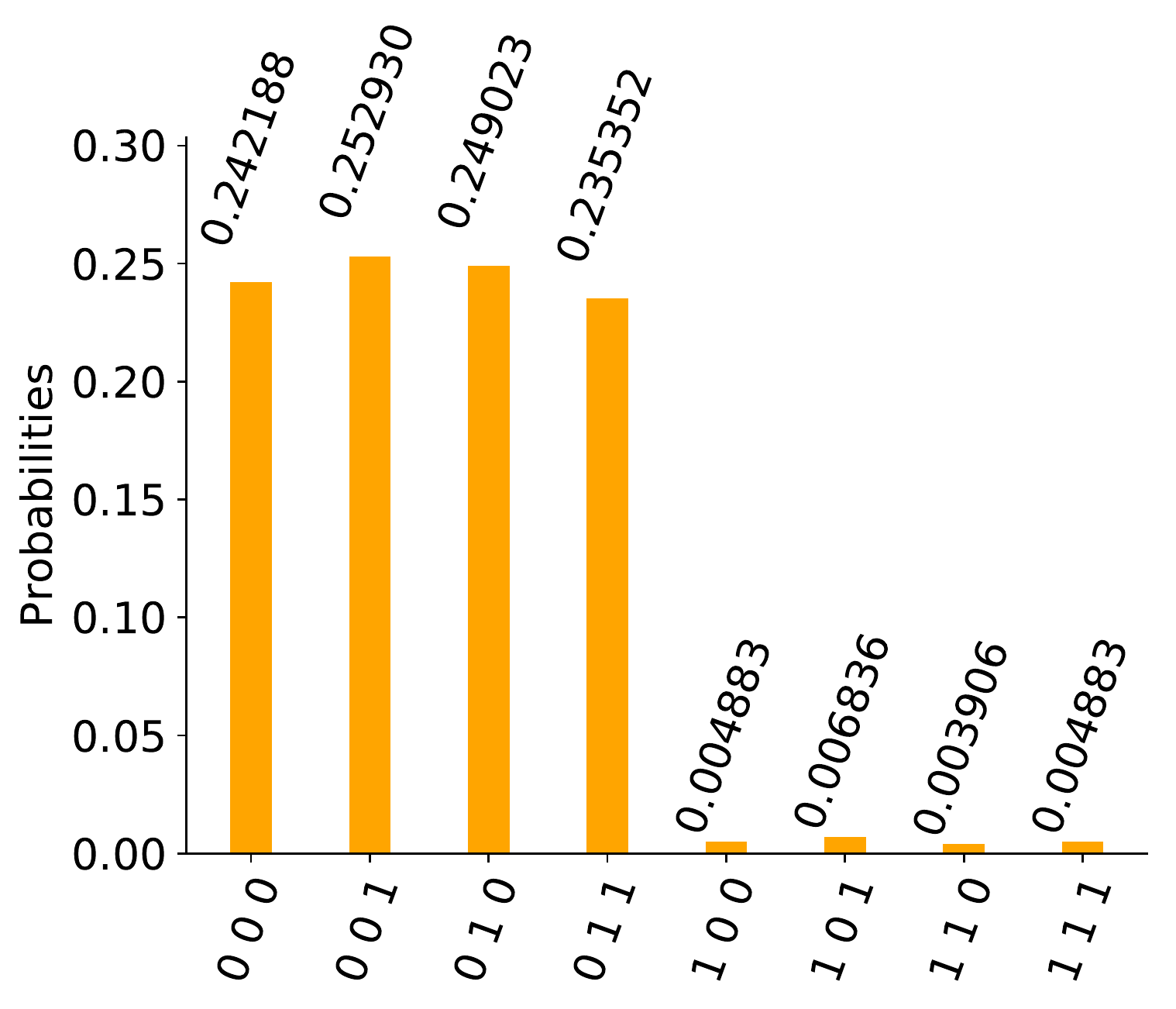}
\caption{Coupling map 2.} \label{fig:5}
\end{subfigure}\hspace*{\fill}
\begin{subfigure}{0.33\textwidth}
\centering
\includegraphics[width=\linewidth,height=3cm]{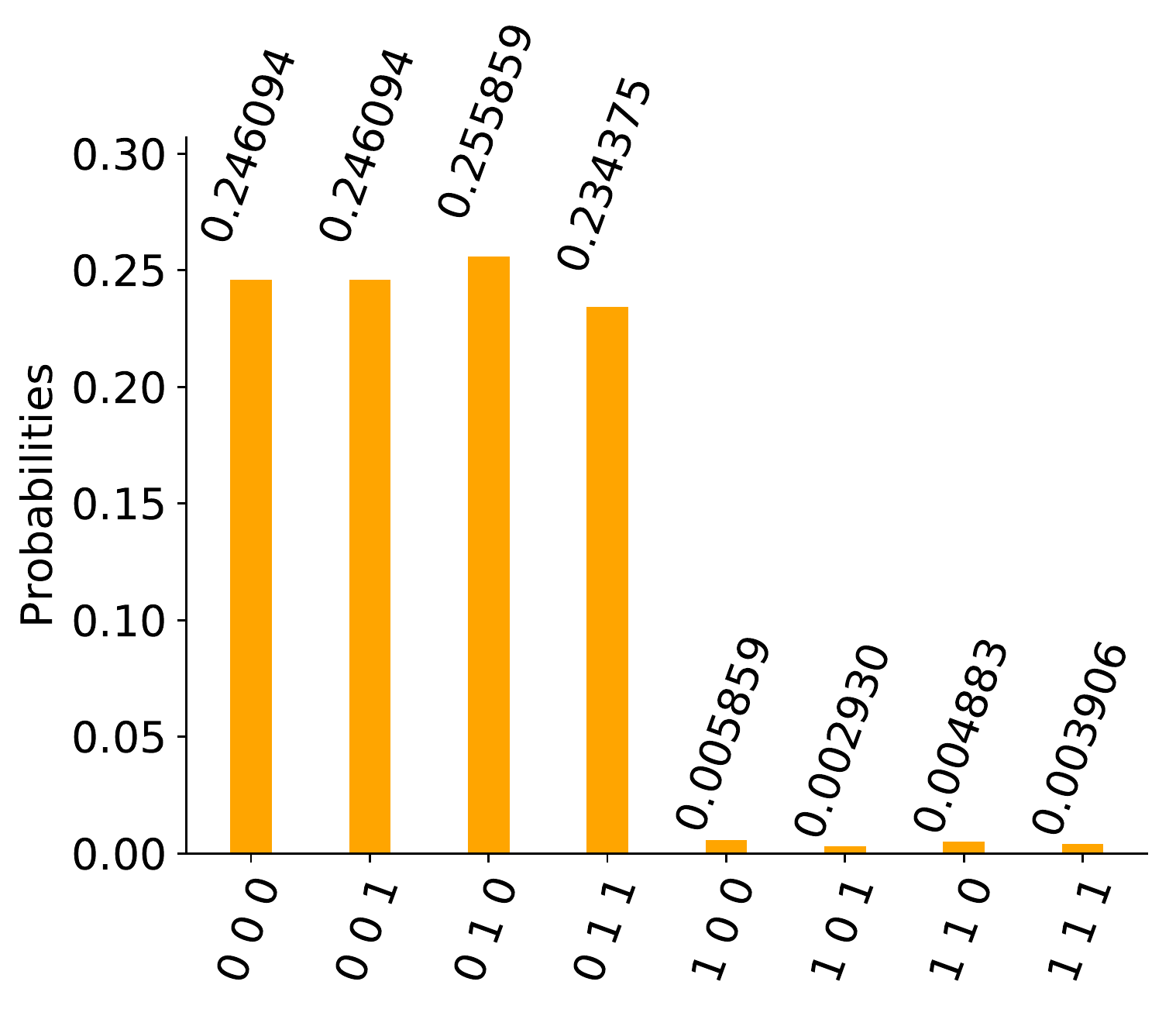}
\caption{Coupling map 3.} \label{fig:6}
\end{subfigure}
\caption{5-qubits coupling map for quantum teleportation.}\label{figure}
\end{figure}

Our results show that an quantum coupling map algorithm based on tabu search can produce the same behaviour as not using any mapping or by setting a predefined map. However, by restricting multi-qubit operations to coupling maps, we decrease the state decoherence\footnote{A process that separates states so that they can no longer interfere.} in qubits. 

\section{Conclusion and future work}
\label{sec:conclusion}

Quantum computers calculate multiple functions at once, exponentially increasing processing speed with each added qubit. \cite{34, 32} A quantum computer uses quantum states and dynamics of particles to store and process information. Transistors have continually decreasing the size, and doubled the power of computers. When technologies reaches the scale of atoms, quantum effects can disrupt its operations due to effects as tunneling and entanglement.

QTS allows us to discover solutions efficiently, where the entangled state can solve high-dependency problems, by using fewer evaluation to determine the global optimum, it also increases the search speed and probes to be a quantum procedure to escape from local optima. In the future we intend to validate our results by comparing different circuits, and by displaying all the results in a descending order instead of just the best one. Also, we intend to implement the algorithm in a D-Wave machine and compare with the current approach. D-Wave Systems has been developing their own version of a quantum computer that uses annealing, which is different from the gate model based approaches from IBM. In the D-Wave’s regular (forward) quantum annealing computing approach, we begin with a massive amount of data that must be mapped to an energy space. This mapping process is actually a mathematical strategy, like simulated annealing\footnote{A metaheuristic to approximate global optimization in a large search space.} or parallel tempering\footnote{A super simulated annealing where a system at high temperature can supply new local optimizers to a system at low temperature, allowing tunneling and improving convergence to a global optimum.}. This procedure allows to map a highly quantum mechanical state as a superposition of all the potential solutions. Then the D-Wave machine slowly fades the quantum state and quantum tunneling\footnote{Effect where a particle crosses through a classically forbidden potential energy barrier.}, superposition occurs, and entanglement and coherence manages interactions. As the quantum mechanical wave function laid across possible solutions, it shows the solution sets that are most accurate. Nowadays, D-Wave makes quantum leap with reverse annealing. It is now possible to give results from classical algorithms to the quantum annealer and work backwards. The system starts with the classical state, then falls back in the annealing process introducing quantum dynamics, rather than coming from a superposition of all possible states (candidate states) with equal weights, which is a more similar approach to our implementation.

\bibliographystyle{plainnat}
\bibliography{references}

\end{document}